\documentclass[preprint2]{aastex}
\usepackage{graphicx}
\usepackage{natbib}
\bibpunct{(}{)}{;}{a}{}{,}

\shorttitle{X-ray afterglow light curves : toward standard candle ?}
\shortauthors{Gendre et al.}

\begin{document}
   \title{X-ray afterglow light curves : toward standard candle ?}

\author{B. Gendre\altaffilmark{1,2,3}, A. Galli\altaffilmark{4}}
\affil{IASF-Roma, via fosso del cavaliere 100, 00133, Roma, Italy}

\and

\author{M. Bo\"er}
\affil{Observatoire de Haute Provence (CNRS), 04870 Saint Michel l'Observatoire, France}

\altaffiltext{1}{Bicocca University, Milano, Italy}
\altaffiltext{2}{Laboratoire d'Astrophysique de Marseille/CNRS/Universit\'e de Provence, BP 8, 13376 Marseille Cedex 12, France}
\altaffiltext{3}{bruce.gendre@oamp.fr}
\altaffiltext{4}{INFN-Trieste, Trieste, Italy}

   \date{Received ---; accepted ---}
 
\begin{abstract}
We investigate the clustering of afterglow light curves observed at X-ray and  optical wavelengths. We have constructed a sample of 61 bursts with known distance and X-ray afterglow. This sample includes bursts observed by BeppoSAX, XMM-Newton, Chandra, and SWIFT. We correct the light curves for cosmological effects and compare the observed X-ray fluxes one day after the burst. We check for correlations between the observed flux and the burst spectral and temporal properties. We confirm the previous result of Bo\"er and Gendre (2000) that X-ray afterglow light curves cluster in luminosity, even when we consider the last SWIFT data. We observe this clustering only for the afterglow light curves; the inclusion of prompt-related data broaden the distribution. A similar clustering is observed for the optical light curves; GRB sources can be divided in three classes, namely optical and X-ray {\it bright} afterglows, optical and X-ray {\it dim} ones, and optically {\it bright} -X-ray {\it dim} ones. We argue that this clustering is related to the fireball total energy, the external medium density, the fraction of fireball energy going in relativistic electrons and magnetic fields. These parameters can be either  fixed to a standard value, or correlated.  We finally propose a method for the estimation of the GRB source redshift based on the observed X-ray flux one day after the burst and optical properties. Using this method, we compute a redshift of 1.4 $\pm$ 0.2 for \object{GRB 980519} and of $1.9 \pm 0.3$ for \object{GRB 040827}. We tested this method on three recently detected SWIFT GRBs with known redshift, and found it in good agreement with the reported distance from optical spectroscopy.
  \end{abstract}


   \maketitle
%

\section{Introduction}

Long Gamma-Ray Bursts (GRBs) are linked with the death of massive stars \citep[for a review, see][]{mes06}. Their association with supernovae \citep[e.g.][]{hjo03, sta03} and the fact that these events are at cosmological distance \citep{met97} make them interesting for studies of cosmology in the redshift range 1-15. However, while the GRB features a prompt emission, usually seen in gamma-ray only, and an afterglow seen at all wavelengths, only the former emission has been considered for cosmological studies yet. One of the first attempt to do so, based on the $E_p-E_{iso}$ correlation found by \citet{ama02}, was done by \citet{ghi03}, who constrained the $\Omega_m - \Omega_\Lambda$ parameters. While not very constraining, their findings were compatible with previous tests made with supernova samples. The use of GRB for cosmological studies, specially at high redshift where their detection at optical/IR wavelength is difficult, needs to build a robust indicator of their distance, whenever possible based on their intrinsic properties.

The afterglow emission is less studied from a cosmological point of view because of its diversity. However, there were hints of standardization of the X-ray afterglows. The first attempt was done by \citet{boe00}, who found evidences for clustering in the X-ray light curves of BeppoSAX afterglows. This clustering was confirmed later by \citet{gen05} who extended the sample to the XMM-Newton and Chandra data. In the following we will refer to these articles respectively as paper I and II. In addition, \citet{kou04} showed that supernova and GRB light curves had similar behaviors and were converging with time toward a similar luminosity. In paper I we tried also to check if optical light curves were also   clustered: this attempt failed due to the poor knowledge of the intrinsic absorption in the burst host galaxy at that time. This study was completed by \citet{nar06} and \citet{lia06} who found independently that optical afterglows were also clustered in luminosity.

The high detection rate and throughput of SWIFT, which provide a large sample of X-ray observations and rapid, accurate localization, enabling a redshift estimation by optical/IR telescopes, allowed us to increase dramatically our sample. With this larger sample, we have tried to derive a method for estimating the burst redshift from the X-ray light curve \citep{gen06}. However, before to apply this method it is necessary to understand the nature of these two groups, and to determine a consistent way to derive to which group belongs each burst. This is the purpose of this article. In Section \ref{sec_ana} we present our sample and the data analysis we performed. We discuss the X-ray clustering in Section \ref{sec_result}. We compare the X-ray and optical clusterings in Section \ref{sec_discu}. In section \ref{sec_redshift} we present our method of GRB source redshift estimation from the X-ray afterglow light curve. We finally test this method on several GRBs detected recently by SWIFT, and we propose an estimation of the redshift for two previously detected GRBs of unknown distance.

\section{X-ray afterglow sample and analysis}
\label{sec_ana}

Our sample of GRBs with known redshift and X-ray afterglow observations is listed in Table \ref{table1}. It includes all afterglows observed by BeppoSAX, XMM-Newton and Chandra, which data were retrieved from \citet{dep06} and \citet{gen05b}. As for SWIFT observations, we browsed the SWIFT archive web page and selected all bursts detected prior to the 1st of August 2006 with a measured redshift and a $t_{90}$ larger than 2.0 seconds (in order to exclude short bursts). From this sample of SWIFT bursts, we further excluded \object{GRB 060123} due to data processing errors. We removed the flaring parts of the light curve when applicable (this totally removed from the sample the data of \object{GRB 050904}, which is thus not listed in Table \ref{table1}). We corrected the flux light curves for distance effects as in Paper I and II : we apply a k-correction using the measured spectral index of each afterglow, assuming a flat universe with $\Omega_m = 0.3$, and correcting for the time dilation effect by computing the luminosity at a time $t$ using the observed flux at time $t \times (1+z)$. This is very important for the correction accuracy, as discussed in Sec. \ref{sec_discu}. We restricted the light curves to the 2.0$-$10.0 keV X-ray band, where the absorption is negligible. This allowed us to neglect any other corrections for absorption by the ISM. We do not take into account any beaming effect due to a possible jet structure.

As we done previously, instead of using a luminosity light curve, we express all light curves in flux units at a given distance (like the optical absolute magnitude). This allows to reduce the uncertainties on the correction. For consistency with paper I and II, we fix the redshift to $z=1$.

\begin{deluxetable}{cccccccccc}
\tabletypesize{\footnotesize}
\tablecaption{The burst sample used in this paper. We indicate for each burst the satellite that observed the X-ray afterglow, its redshift, spectral and decay index, its group assignation (see text) in X-rays and in optical \citep[extracted from][]{nar06, lia06}, and the values of $E_{iso}$ and $E_p$ listed in \citet{ama06}. For the bursts detected by SWIFT we also indicate the $T_a$ values listed in \citet{wil07} corresponding to the afterglow start. The spectral and decay index information are extracted from \citet[][ BeppoSAX]{dep06}, \citet[][ XMM-Newton, Chandra]{gen05b}, and \citet[][ SWIFT]{wil07}.\label{table1}}
\tablehead{\colhead{Burst name} &\colhead{X-ray} &\colhead{redshift} &\colhead{Decay} &\colhead{Spectral} &\colhead{X-ray} &\colhead{Optical} &\colhead{$E_{iso}$} &\colhead{$E_p$} &\colhead{log($T_a$)}\\
\colhead{     }  &\colhead{satellite} &\colhead{       }  &\colhead{index} & \colhead{index}   &\colhead{group} &\colhead{group}   &\colhead{($10^{52}$ erg)}&\colhead{(keV)}& \colhead{(s)}}
\startdata
\object{GRB 970228} &  BeppoSAX   & 0.695    & $1.3 \pm 0.2$      & $1.0 \pm 0.3$  &  II   &  I      & $1.84\pm0.14$ & $195\pm64 $ & ---\\
\object{GRB 970508} &  BeppoSAX   & 0.835    & $0.8 \pm 0.2$      & $1.4 \pm 0.3$  &  II   &  II ?   & $0.71\pm0.15$ & $145\pm43 $ & ---\\
\object{GRB 971214} &  BeppoSAX   & 3.42     & $1.0 \pm 0.3$      & $1.1 \pm 0.4$  &  I    &  I      & $24 \pm 3   $ & $685\pm133$ & ---\\
\object{GRB 980425} &  BeppoSAX   & 0.0085   & $0.10\pm0.06$      &    (1.1)       &  III  &  ---    &$10^{-4}\pm2\times10^{-5}$ &$55\pm21$& ---\\
\object{GRB 980613} &  BeppoSAX   & 1.096    & $1.5^{+1.9}_{-0.9}$&    (1.1)       &  II   &  II     & $0.68\pm0.11$ & $194\pm89 $ & ---\\
\object{GRB 980703} &  BeppoSAX   & 0.966    & $1.1^{+0.4}_{-0.3}$& $1.7 \pm 0.3$  &  II   &  I      & $8.3 \pm 0.8$ & $503\pm64 $ & ---\\
\object{GRB 990123} &  BeppoSAX   & 1.60     & $1.45\pm0.06$      &$0.99\pm0.05$   &  I    &  I      & $266\pm43   $ & $1724\pm466$& ---\\
\object{GRB 990510} &  BeppoSAX   & 1.619    & $1.4 \pm 0.1$      & $1.17\pm0.09$  &  I    &  I      & $20 \pm 3   $ & $423\pm42 $ & ---\\
\object{GRB 991216} &  Chandra    & 1.02     & $ (0.9)     $      & $0.7 \pm 0.2$  &  I    &  I      & $78 \pm 8   $ & $648\pm134$ & ---\\
\object{GRB 000210} &  BeppoSAX   & 0.846    & $1.38\pm0.03$      & $0.9 \pm 0.2$  &  II   &  ---    & $17.3\pm1.9 $ & $753\pm26 $ & ---\\
\object{GRB 000214} &  BeppoSAX   & 0.42     & $0.7 \pm 0.4$      & $1.0 \pm 0.3$  &  II   &  ---    & $0.93\pm0.03$ & $>117     $ & ---\\
\object{GRB 000926} &  BeppoSAX   & 2.066    & $1.9 \pm 0.4$      & $0.7 \pm 0.2$  &  I    &  I      & $31.4\pm6.8 $ & $310\pm20 $ & ---\\
\object{GRB 010222} &  BeppoSAX   & 1.477    & $1.35\pm0.06$      & $1.00\pm0.06$  &  I    &  I      & $94 \pm 10  $ & $766\pm30 $ & ---\\
\object{GRB 011121} &  BeppoSAX   & 0.36     & $1.30\pm0.03$      &     (1.1)      &  II   &  II     & $9.9 \pm 2.2$ & $793\pm533$ & ---\\
\object{GRB 011211} &  XMM        & 2.14     & $2.1 \pm 0.3$      & $1.3 \pm 0.3$  &  II   &  I      & $6.3\pm0.7  $ & $186\pm24 $ & ---\\
\object{GRB 020405} &  Chandra    & 0.69     & $1.6 \pm 0.9$      & $1.0 \pm 0.2$  &  II   &  I      & $12.8\pm1.5 $ & $612\pm122$ & ---\\
\object{GRB 020813} &  Chandra    & 1.25     & $1.4 \pm 0.2$      & $0.83\pm0.06$  &  I    &  I      & $76\pm19    $ & $590\pm151$ & ---\\
\object{GRB 021004} &  Chandra    & 2.3      & $1.2 \pm 0.2$      & $1.01\pm0.07$  &  I    &  I      & $3.8\pm0.5  $ & $266\pm117$ & ---\\
\object{GRB 030226} &  Chandra    & 1.98     & $2.7 \pm 1.6$      & $0.9 \pm 0.3$  &  II   &  I      & $14\pm1.5   $ & $289\pm66 $ & ---\\
\object{GRB 030328} &  Chandra    & 1.52     & $1.6 \pm 0.3$      & $1.1 \pm 0.2$  &  II   &  I      & $43 \pm 4   $ & $328\pm55 $ & ---\\
\object{GRB 030329} &  XMM        & 0.168    & $  (2)      $      & $1.0 \pm 0.2$  &  II   &  I      & $1.7 \pm 0.3$ & $100\pm23 $ & ---\\
\object{GRB 031203} &  XMM        & 0.105    & $0.5 \pm 0.1$      & $0.8 \pm 0.2$  &  III  &  ---    &$0.010\pm0.004$& $158\pm51 $ & ---\\
\object{GRB 050126} &  SWIFT      & 1.29     & $1.1^{+0.6}_{-0.5}$& $0.7 \pm 0.7$  &  II   &  ---    &    ---        &  ---        & $2.34^{+3.30}_{-1.00}$\\
\object{GRB 050223} &  SWIFT      & 0.5915   &                    & $1.4 \pm 0.7$  &  III  &  ---    &    ---        &  ---        & \\
\object{GRB 050315} &  SWIFT      & 1.949    & $0.7^{+0.2}_{-0.1}$& $0.91\pm0.09$  &  I    &  I      & $4.9 \pm1.5 $ & $<89      $ & $4.39 \pm 0.27$       \\
\object{GRB 050319} &  SWIFT      & 3.24     & $1.4^{+0.5}_{-0.4}$& $0.96\pm0.09$  &  I    &  I      &    ---        &   ---       & $4.67 \pm 0.27$       \\
\object{GRB 050401} &  SWIFT      & 2.90     & $1.5^{+0.5}_{-0.2}$& $1.0 \pm 0.3$  &  I    &  I      & $41 \pm 8   $ & $467\pm110$ & $3.87^{+0.24}_{-0.18}$\\
\object{GRB 050406} &  SWIFT      & 2.44     & $1.0^{+4.1}_{-1.0}$& $1.0 \pm 0.4$  &  II   &  ---    &    ---        &   ---       & $2.62^{+9.6}_{-1.0}$  \\
\object{GRB 050416A}&  SWIFT      & 0.6535   & $ 0.85 \pm 0.05$   & $1.1 \pm 0.1$  &  II   &  ---    & $0.12\pm0.02$ & $25.1\pm4.2$& $3.19^{+0.3}_{-0.5}$  \\
\object{GRB 050505} &  SWIFT      & 4.27     & $1.5^{+0.6}_{-0.2}$& $1.0 \pm 0.1$  &  I    &  I      &    ---        &  ---        & $4.39^{+0.48}_{-0.24}$\\
\object{GRB 050525A}&  SWIFT      & 0.606    & $ 1.4 \pm 0.1$     & $1.1 \pm 0.4$  &  II   &  I      & $3.39\pm0.17$ & $127\pm10 $ & $2.92^{+0.14}_{-1.23}$\\
\object{GRB 050603} &  SWIFT      & 2.821    & $1.8^{+0.5}_{-0.3}$& $0.7 \pm 0.1$  &  I    &  I      & $70 \pm 5$    & $1333\pm107$& $4.83^{+0.30}_{-1.14}$\\
\object{GRB 050730} &  SWIFT      & 3.968    & $2.7^{+0.3}_{-0.2}$& $0.62 \pm 0.05$&  I    &  I      &    ---        &  ---        & $4.13 \pm 0.08 $      \\
\object{GRB 050802} &  SWIFT      & 1.71     & $1.7 \pm 0.2$      & $0.81 \pm 0.09$&  II   &  ---    &    ---        &  ---        & $3.96 \pm 0.10 $      \\
\object{GRB 050814} &  SWIFT      & 5.3      & $0.8 \pm 0.4$      & $0.7 \pm 0.1$  &  I    &  ---    &    ---        &  ---        & $3.93^{+0.56}_{-0.70}$\\
\object{GRB 050824} &  SWIFT      & 0.83     & $0.8^{+0.5}_{-0.2}$& $0.8 \pm 0.2$  &  II   &  II     &$0.130\pm0.029$& $<23      $ & $4.82^{+1.84}_{-0.54}$\\
\object{GRB 050826} &  SWIFT      & 0.297    & $1.13 \pm 0.04$    & $1.1 \pm 0.4$  &  III  &  ---    &    ---        &  ---        & ---\\
\object{GRB 050908} &  SWIFT      & 3.344    & $1.1^{+0.3}_{-0.4}$&                &  II   &  ---    &    ---        &  ---        & $3.31^{+0.86}_{-1.29}$\\
\object{GRB 050922C}&  SWIFT      & 2.198    & $1.26 \pm 0.04$    & $1.3 \pm 0.2$  &  II   &  I      & $6.1 \pm 2.0$ & $415\pm111$ & $2.58 \pm 0.11 $      \\
\object{GRB 051016B}&  SWIFT      & 0.9364   & $0.7^{+0.2}_{-0.3}$& $0.9 \pm 0.2$  &  II   &  ---    &    ---        &  ---        & $3.51^{+0.64}_{-0.81}$\\
\object{GRB 051109A}&  SWIFT      & 2.346    & $1.25 \pm 0.07$    & $1.0 \pm 0.2$  &  I    &  ---    &    ---        &  ---        & $3.93^{+0.15}_{-0.23}$\\
\object{GRB 051109B}&  SWIFT      & 0.080    & $1.1 \pm 0.3$      & $0.7 \pm 0.4$  &  III  &  ---    &    ---        &  ---        & $3.67^{+0.27}_{-0.40}$\\
\object{GRB 051111} &  SWIFT      & 1.549    & $1.56 \pm 0.02$    & $1.1 \pm 0.4$  &  II   &  I      &    ---        &  ---        & ---\\
\object{GRB 060108} &  SWIFT      & 2.03     & $1.3^{+0.3}_{-0.2}$& $1.0 \pm 0.3$  &  II   &  ---    &    ---        &  ---        & $4.40\pm0.2$          \\
\object{GRB 060115} &  SWIFT      & 3.53     & $1.0^{+0.2}_{-0.3}$& $1.3 \pm 0.3$  &  I    &  ---    &    ---        &  ---        & $3.86^{+1.61}_{-1.00}$\\
\object{GRB 060206} &  SWIFT      & 4.048    &$1.24^{+0.05}_{-0.06}$&$0.8 \pm 0.2$ &  I    &  ---    &    ---        &  ---        & $3.86^{+0.14}_{-0.18}$\\
\object{GRB 060210} &  SWIFT      & 3.91     & $1.8^{+1.1}_{-0.3}$& $1.00 \pm 0.09$&  I    &  ---    &    ---        &  ---        & $4.46^{+0.33}_{-0.29}$\\
\object{GRB 060218} &  SWIFT      & 0.033    & $1.3^{+1.1}_{-0.6}$& $0.51\pm0.05$  &  III  &  ---    &    ---        &  ---        & $5.01^{+0.63}_{-0.48}$\\
\object{GRB 060223A}&  SWIFT      & 4.41     & $1.3 \pm 0.2$      & $0.9 \pm 0.3$  &  II   &  ---    &    ---        &  ---        & $2.73^{+0.32}_{-0.36}$\\
\object{GRB 060418} &  SWIFT      & 1.489    & $1.5 \pm 0.2$      & $0.8 \pm 0.9$  &  II   &  ---    &    ---        &  ---        & $3.44^{+0.14}_{-0.18}$\\
\object{GRB 060510B}&  SWIFT      & 4.9      & $1.0^{+0.5}_{-0.3}$& $1.7 \pm 0.4$  &  I    &  ---    &    ---        &  ---        & $4.55^{+0.60}_{-0.54}$\\
\object{GRB 060512} &  SWIFT      & 0.4428   & $1.2 \pm 0.2$      & $0.9 \pm 0.2$  &  III  &  ---    &    ---        &  ---        & $3.85^{+0.34}_{-1.0}$ \\
\object{GRB 060522} &  SWIFT      & 5.11     & $1.0^{+0.2}_{-0.1}$& $1.1 \pm 0.2$  &  I    &  ---    &    ---        &  ---        & $2.86^{+0.46}_{-0.25}$\\
\object{GRB 060526} &  SWIFT      & 3.21     & $1.1^{+0.3}_{-0.2}$& $0.7 \pm 0.3$  &  I    &  ---    &    ---        &  ---        & $3.84^{+0.47}_{-0.33}$\\
\object{GRB 060604} &  SWIFT      & 2.68     & $1.2 \pm 0.2$      & $1.3 \pm 0.2$  &  ?    &  ---    &    ---        &  ---        & $4.55 \pm 0.29 $      \\
\object{GRB 060605} &  SWIFT      & 3.78     & $2.0^{+0.4}_{-0.3}$& $1.0 \pm 0.1$  &  I ?  &  ---    &    ---        &  ---        & $4.16 \pm 0.13 $      \\
\object{GRB 060607A}&  SWIFT      & 3.082    & $8.5^{+8.5}_{-0.4}$& $0.86 \pm 0.09$&  II   &  ---    &    ---        &  ---        & $4.75 \pm 0.03 $      \\
\object{GRB 060614} &  SWIFT      & 0.125    & $2.0^{+0.3}_{-0.2}$& $0.8 \pm 0.2$  &  III  &  ---    &    ---        &  ---        & $5.00 \pm 0.09 $      \\
\object{GRB 060707} &  SWIFT      & 3.43     & $0.83 \pm 0.08$    & $1.2 \pm 0.5$  &  I    &  ---    &    ---        &  ---        & $3.58^{+0.47}_{-0.71}$\\
\object{GRB 060714} &  SWIFT      & 2.71     &$1.25^{+0.09}_{-0.07}$&$1.4 \pm 0.3$ &  I    &  ---    &    ---        &  ---        & $3.20^{+0.21}_{-0.41}$\\
\object{GRB 060729} &  SWIFT      & 0.54     & $1.32 \pm 0.05$    & $1.39 \pm 0.05$&  II   &  ---    &    ---        &  ---        & $5.11 \pm 0.04 $      \\

\enddata
\end{deluxetable}

\section{Results}
\label{sec_result}
\subsection{Is the clustering still apparent?}

\begin{figure}
  \includegraphics[height=.27\textheight]{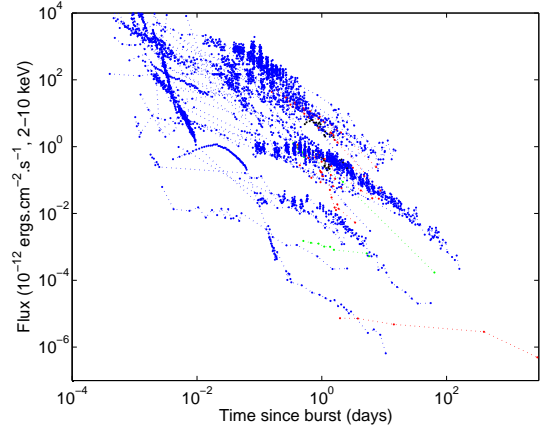}
  \caption{\label{Fig1} The light curves of our sample, corrected for distance effects. We indicate SWIFT, XMM-Newton, Chandra and BeppoSAX bursts with blue, green, black and red diamonds respectively. 
  The error bars are not plotted for clarity. See electronic version for colors.}
\end{figure}

 \begin{figure*}
   \centering
   \includegraphics[width=8.4cm]{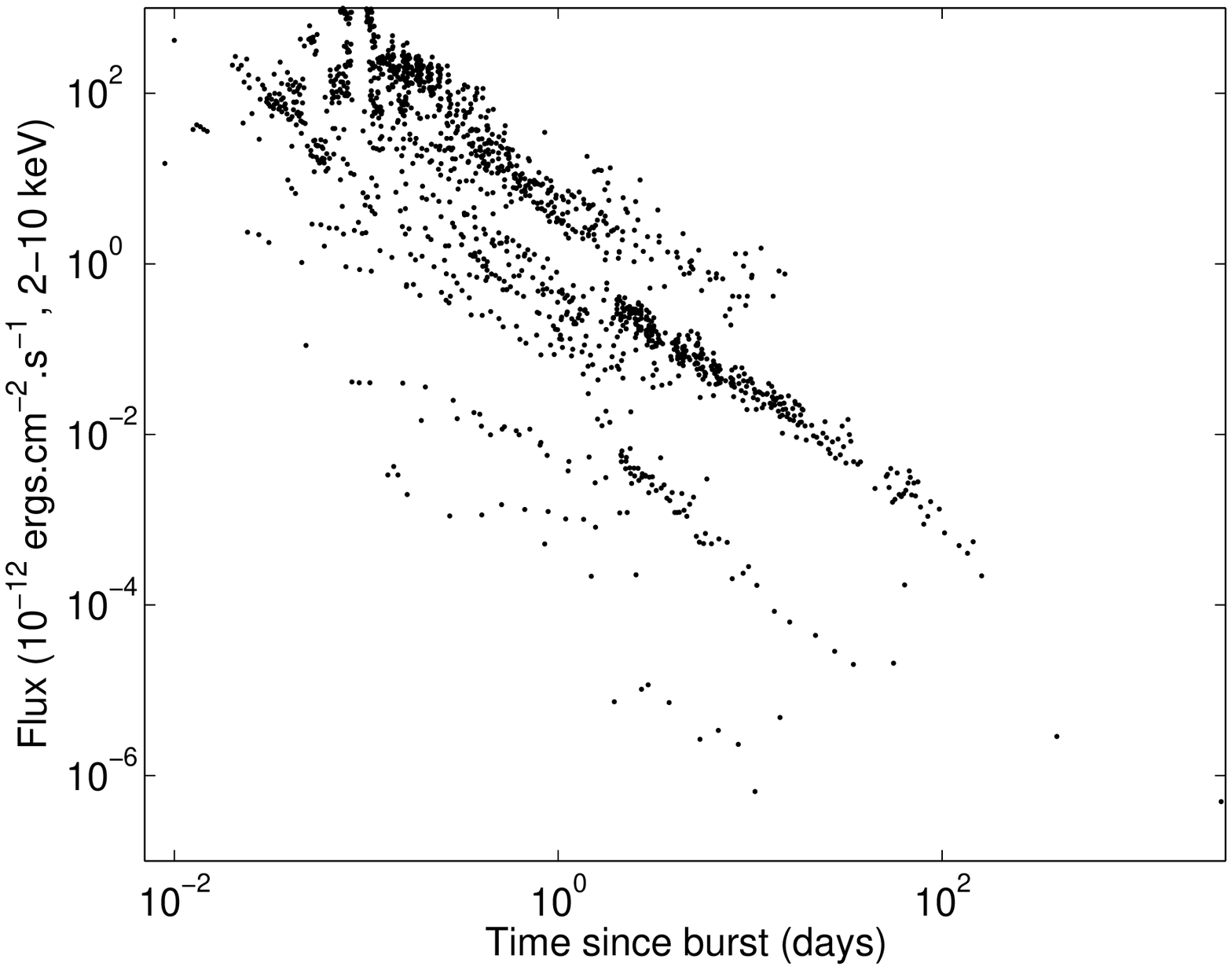}
   \includegraphics[width=8cm]{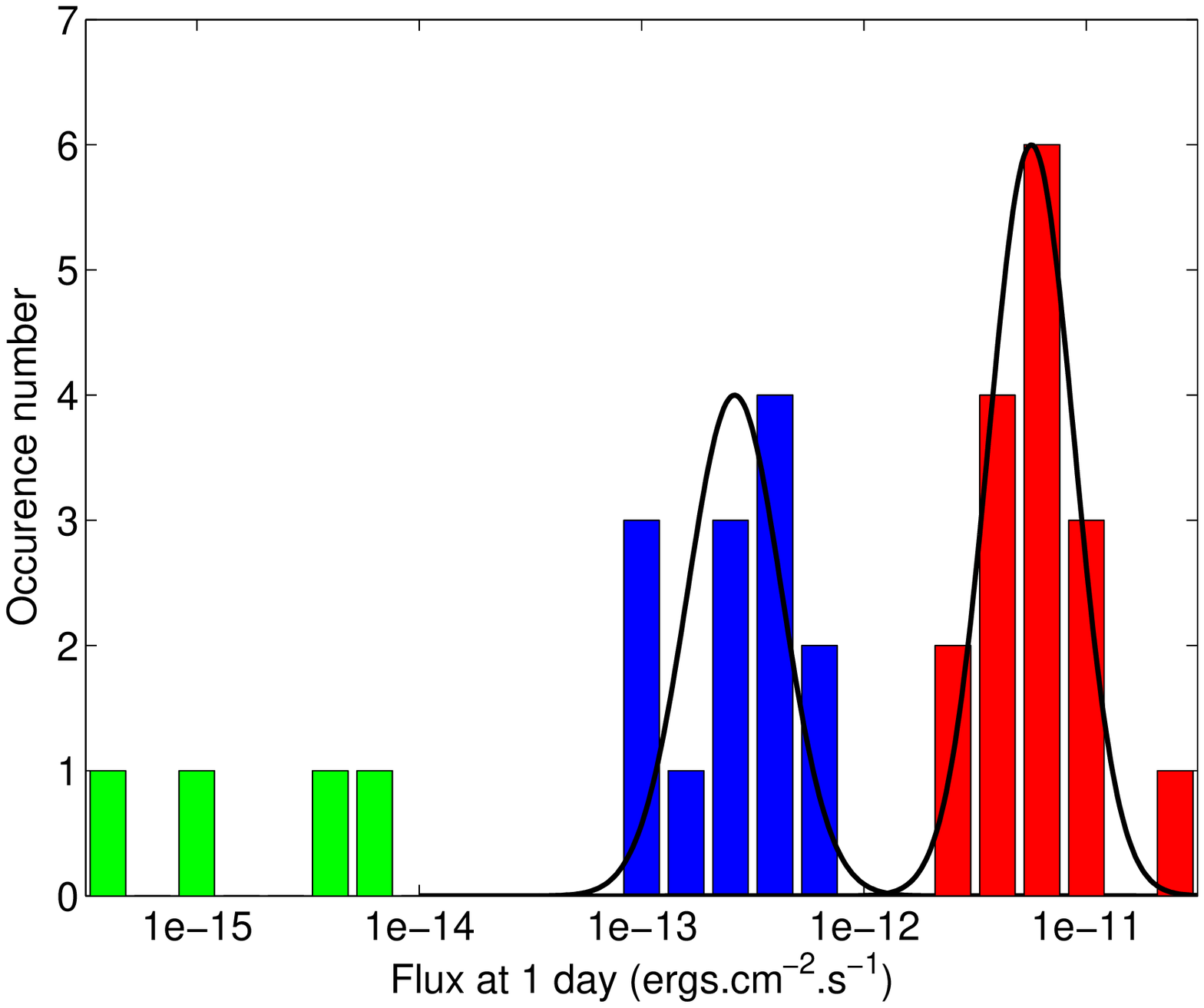}
      \caption{Left: the corrected light curves of our sample, using only data taken after the plateau phase for SWIFT bursts (see text). The groups reported in papers I and II are now clearly seen. Right: the distribution of fluxes observed at one day after the burst. The solid lines are the best Gaussian fit distribution of the group fluxes. In the following figures, red refers to bright events, blue to intermediate luminosity events and green to under-luminous events. See electronic version for colors.
              }
         \label{fig2}
   \end{figure*}

Figure \ref{Fig1} presents the light curves corrected for distance effects. As one can clearly see, no clustering is observed in the early part of the light curves. The addition of SWIFT bursts seems to have blurred out the clustering properties observed in paper I and II. 
If we assume that SWIFT is observing GRBs similar to those observed previously, the only difference that could explain this result is the addition of the early time X-ray light curves, since previous observations from BeppoSAX, XMM, Chandra where performed usually at least 6h after the GRB alert. 

\subsection{Influence of the prompt emission}

SWIFT has shown that the X-ray light curve is typically composed by a first steep power-law segment, associated with the tail of prompt emission, followed by a flat plateau, a second steepening (which is the segment usually seen before SWIFT), and a possible late steepening related to the jet aperture \citep{nou05, obr06, zhe06}. \citet{wil07} have interpreted this behavior with a two-components model. According to them, the plateau phase marks the transition between a first component, ascribed to the prompt emission, and a second one which nature is less clear and develops in the afterglow. We have retrieved for SWIFT bursts the values of the end-time $T_a$ of the plateau phase, and excluded all SWIFT data taken before this time. The results are displayed in Fig. \ref{fig2}.

Excluding these data, the two groups reported in papers I and II become apparent. We note however still some dispersion during the first part of the light curves. We interpret this as a consequence on the error (sometime large) on the $T_a$ measurement. Indeed, restricting the sample to data collected after $T_a+\sigma_{Ta}$ (i.e. being very conservative on the $T_a$ value) strongly reduce this dispersion, at the price of a drastic reduction of the sample size. In the following, we will use all data taken after $T_a$ (i.e. a less conservative hypothesis) and do not discuss the effects seen in the early part of the light curves. We refer to the bright group as {\it group I} and the dimmer one as {\it group II}, as in paper II. They are clustering with a mean flux of $7.0 \times 10^{-12}$ erg s$^{-1}$ cm$^{-2}$ and $3.1 \times 10^{-13}$ erg s$^{-1}$ cm$^{-2}$ for groups {\it I} and {\it II} respectively, assuming a common redshift of unity (see above) and at a time of 1 day. Some bursts do not follow this relation : \object{GRB 980425}, \object{GRB 031203}, \object{GRB 050223}, \object{GRB050826}, \object{GRB 051109B}, \object{GRB 060218}, \object{GRB 060512}, and \object{GRB 060614}. They are all nearby events, the most distant, \object{GRB 060512}, being at $z = 0.4428$ \citep{blo06}. Note however that \object{GRB 030329} is also at low redshift ($z = 0.168$) but belongs to {\it group II}.  We consider that these low luminosity events form a specific group, referred as {\it group III} in the following. We finally observe two peculiar events. \object{GRB 060605} belongs to {\it group I} but, due to its steep decay, changes to {\it group II} later on. We note that, contrary to the other bursts, this event is not compatible with an ISM or wind medium, but is compatible with a jet effect. This would imply that the Bo\"er \& Gendre relation is valid {\it before} jet effects become apparent. \object{GRB 060604} is located between {\it group I} and {\it group II}. On 61 events, this is the only one lying within the gap. We note that the light curve of this event presents a strong "noise" (i.e. small scale temporal variations) during the whole observation. A possible explanation to these variations would be that this event presents several X-ray flares (like e.g. \object{GRB 050904}), implying that we do not observe the continuum. Under this assumption, this event would be a {\it group II} event (and should not be considered for further discussion). However, we cannot rule out a possible outliers hypothesis and maintain this event within the sample.

The probability that a power law luminosity distribution (letting the index be a free parameter) represents the observed distribution is, at the maximum, $8.9 \times 10^{-13}$; thus the observed clustering in two groups is very significant. To compute this probability we impose a lower luminosity limit such that the {\it group III} bursts are excluded, because of the selection effects (see Sec. \ref{sec_discu}).

\begin{figure}
   \centering
   \includegraphics[width=8cm]{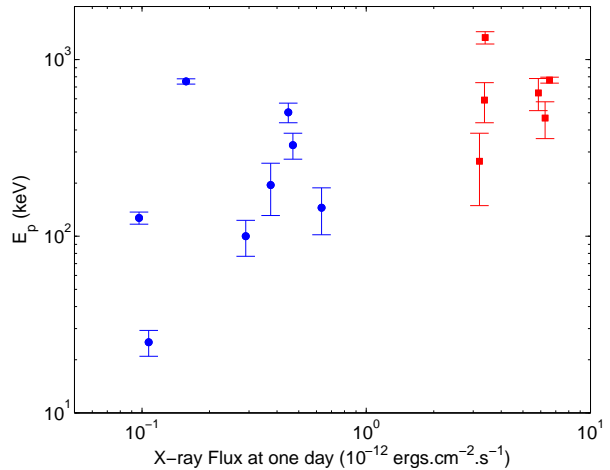}
      \caption{Distribution of intrinsic E$_{p}$ (rest frame) values. Blue circles and red squares represent {\it group II} and {\it group I} bursts respectively. The flux at one day as been computed by interpolating the flux at that date when the burst was observed one day after the burst (see text for discussion about extrapolation effects), thus lowering the number of events available for comparison.
              }
         \label{ep}
   \end{figure}

\begin{figure}
   \centering
   \includegraphics[width=8cm]{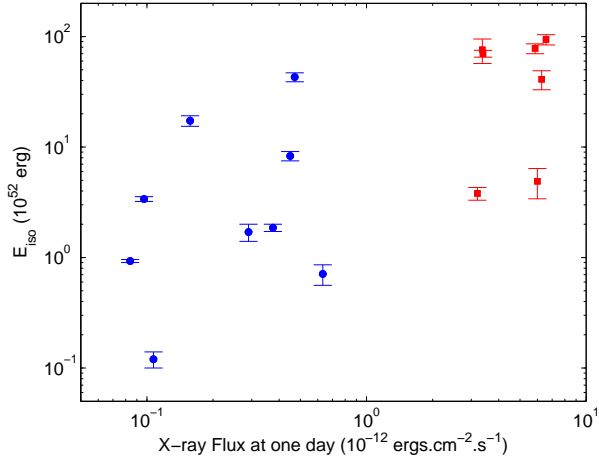}
      \caption{Distribution of the energy emitted during the prompt event assuming an isotropic emission (E$_{iso}$). Symbols are the same as in Fig. \ref{ep}.
              }
         \label{eiso}
   \end{figure}

We checked if this clustering can be related to the isotropic burst energy $E_{iso}$ or the peak energy $E_p$ values. Figure \ref{ep} presents the $E_p$ distribution of bursts versus the X-ray afterglow flux at one day. One can clearly see that the two groups share a similar $E_p$ distribution. This also holds for the $E_{iso}$ distribution (Fig. \ref{eiso}; see also Sec. \ref{sec_discu} for a discussion on the selection effects). Recently, \citet{cha07} reported three classes of GRBs based on their fluence and duration. However, their classification and the results we report here are not correlated. Moreover the early SWIFT X-ray light curves associated with the tail of the prompt emission do not cluster in luminosity. This leads us to consider that the afterglow emission is not correlated with the prompt emission.

\subsection{Influence of the afterglow properties}

  \begin{figure}
   \centering
   \includegraphics[width=8cm]{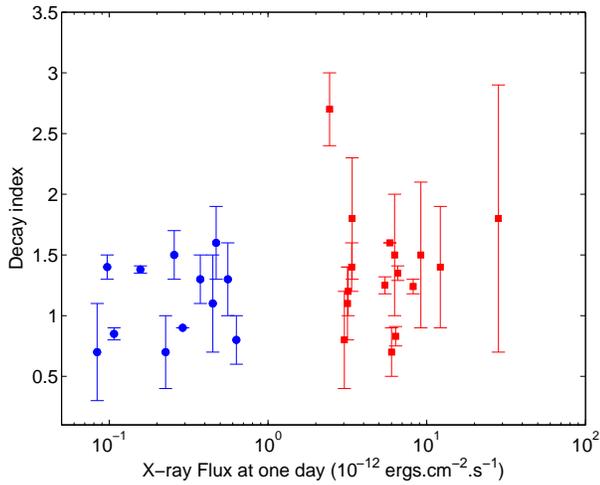}
      \caption{Distribution of the decay index of the groups {\it I} and {\it II}. No differences can be seen. Symbols are the same as in Fig. \ref{ep}.
              }
         \label{decay}
   \end{figure}

\begin{figure}
   \centering
   \includegraphics[width=8cm]{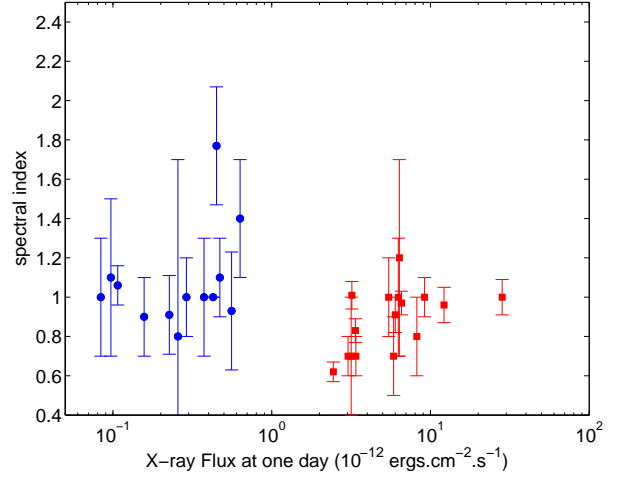}
      \caption{Distribution of the spectral index of the groups {\it I} and {\it II}. No differences can be seen. Symbols are the same as in Fig. \ref{ep}.
              }
         \label{alpha}
   \end{figure}

In paper II, we pointed out a possible segregation of decay indexes : {\it group I} bursts seemed to decay faster than {\it group II} ones. Thanks to the fast monitoring capabilities of SWIFT we can clearly rule out this hypothesis. Figure \ref{decay} shows that there is no difference between the decay index distributions for the two groups. Moreover, SWIFT has added some events to the {\it group I} with a low decay (e.g. \object{GRB050315}). We find the same results when we look at the spectral indexes (Fig. \ref{alpha}). Clearly, the two distributions are similar. In fact, except for two of them, all bursts are compatible with having a common spectral index of $\beta = 1.0 \pm 0.2$ (with the flux $F\propto t^{-\alpha}\nu^{-\beta}$).

\section{Discussion}
\label{sec_discu}

\subsection{Distance effects and bias}

We present all results using a common distance; however, as we are correcting for distance effects, one may wonder if this could bias our findings.

  \begin{figure}
   \centering
   \includegraphics[width=8cm]{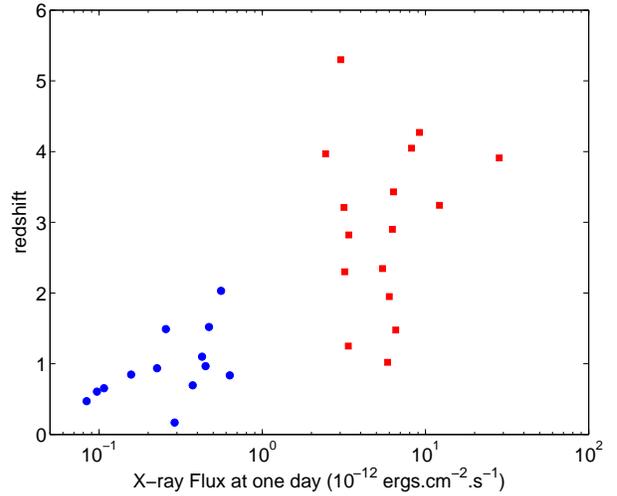}
      \caption{Distribution of the redshifts of the groups {\it I} and {\it II}. Symbols are the same as in Fig. \ref{ep}.
              }
         \label{redshift}
   \end{figure}

Figure \ref{redshift} presents the redshift distribution of our sample. There is a clear trend : low luminosity bursts are nearby while bright ones are more distant. This is even more obvious if we include {\it group III} bursts, which have all a very low redshift. This effect can be partly explained by a selection effect, as low luminosity events cannot be detected at high distance. The lack of dim distant events (located in the upper left corner of Fig. \ref{redshift}) is due to that effect. However, we should also have detected bright nearby events. The nearest {\it group I} burst, (\object{GRB 991216}), is located at $z = 1.02$. While this is already lower than the mean SWIFT GRB redshift \citep[2.7,][]{jac06}, there is a clear deficit of bright sources closer $z \sim 0.5$: whether this is related to the nature, or the intrinsic frequency, of {\it group I} bursts, is still an open question.

Another important effect to take into account is the time dilation and the estimation of the distribution parameters. X-ray light curves are not taken continuously. Before SWIFT, X-ray observations were usually performed as snapshots of 30-60 ksec long \citep{dep06, gen05b}. Because of the time dilation, the start and end of each observation were located at different times, and the estimation of a parameter at a given time needs extrapolation or interpolation from the value observed at another time. However, the decay index can largely vary during the observations as shown by \citet{obr06}; this uncertainty can affect the accuracy of the distribution parameter estimations. This effect can be illustrated by looking at Fig. 6 of \citet{nar06}, which is equivalent to Fig. \ref{fig2} of the present work: the results are clearly different; we observe a strong clustering, while \citet{nar06} see only a trend. However \citet{nar06} extrapolate the burst afterglow light curves, when no data is available at times earlier than 12h in the source rest frame: as an example \object{GRB 020405} is observed from 1.00 day to 1.34 days (source rest frame), \object{GRB 000926} is observed from 0.69 to 4.36 days (idem)\footnote{We stress that this example is given only as illustrative purpose and do not to questions the results of \citet{nar06} at optical wavelengths. As the authors noted, optical light curves have a better temporal coverage and sampling, thus in these case they perform only interpolations.}. The net effect of doing extrapolations instead of interpolations is to broaden the distribution, blurring in turn the clustering reported here. To avoid this effect, we decided, as in paper I and II, to avoid extrapolations and to perform only flux interpolations, when relevant, using the nearest measurements. This is of course at the expense of the sample size, as one can note in Figs. \ref{ep}, \ref{eiso}, \ref{decay}, \ref{alpha}, \ref{redshift} where only half of the total sample is plotted.

\subsection{The optical afterglows}

\citet{kan06}, \citet{nar06} and \citet{lia06} have shown that even optical afterglow light curves display a clustering effect. These authors do not include in their works any of our {\it group III} bursts: for consistency we will restrict the following discussion to groups {\it I} and {\it II}. They found two groups, and we will refer to them as {\it optical group I} (or {\it oI}) and {\it optical group II} (or {\it oII}) for the bright and dim group respectively. We indicate in Table \ref{table1} the optical class of each bursts. Several events of our sample are not included in the work of \citet{nar06} or \citet{lia06}, 	and no optical classification is reported for them. According to \citet{lia06}, one day after the burst (rest frame), the difference in luminosity between groups {\it oI} and {\it oII} is $\sim 26$. At the same time, the difference in luminosity between the X-ray groups is $\sim 24$.

\begin{figure}
   \centering
   \includegraphics[width=8cm]{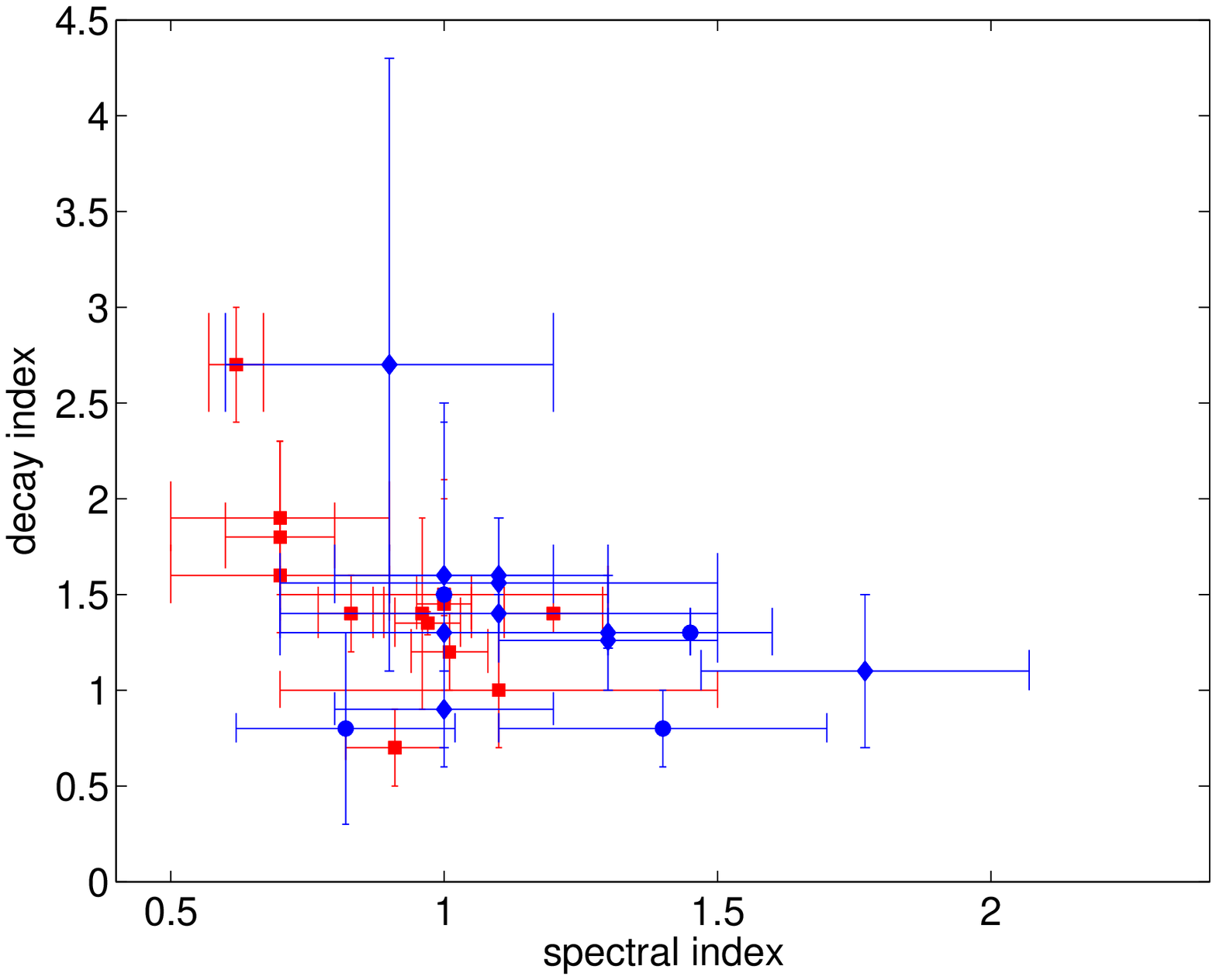}
      \caption{X-ray decay index versus spectral index. In this figure we plot bursts bright both in X-ray and optical (red squares), bursts dim both X-ray and optical (blue circles), and bursts bright in optical and dim in X-ray (blue diamonds); see text for discussion.
              }
         \label{decay-spectral}
   \end{figure}

From Table \ref{table1}, we can see that while an {\it oI} burst can belong to {\it X-ray group I} or {\it X-ray group II}, {\it oII} bursts are dim both at optical and X-ray wavelengths. We present in Fig. \ref{decay-spectral} the distribution of the decay index versus the spectral index of these three burst classes. Again, no clear separation can be observed.

\subsection{The fireball model}

{In this section we investigate further} on the nature of the observed clusterings {in the context of the External Shock model}. \citet{nar06} has indicated that around 1 day for almost all bursts, we have $\nu_m < \nu_{optical} < \nu_c < \nu_X$; thus in the following we will assume this repartition of the characteristic emission frequencies. According to \citet{pan00}, during the slow cooling regime the predicted flux for a fireball expanding in a uniform interstellar medium (ISM) is:

\begin{equation}
\label{eq_ISM_opt}
\begin{array}{lccr}
F_\nu & = & 10^{2.1-1.3p}D_{28}^{-2}\nu_{14.6}^{-(p-1)/2}t_d^{-3(p-1)/4} &\\
       &  & \times E_{53}^{(p+3)/4}n_*^{1/2}\epsilon_{e,~-1}^{p-1}\epsilon_{B,~-4}^{(p+1)/4} mJy &\\
       &  &                                    & (\nu_m < \nu < \nu_c)\\
\end{array}
\end{equation}
       
\begin{equation}
\label{eq_ISM_x}
\begin{array}{lccr}
F_\nu  & =&  10^{2.4-0.8p}D_{28}^{-2}\nu_{14.6}^{-p/2}t_d^{-(3p-2)/4} &\\
       &  &  \times E_{53}^{(p+2)/4}\epsilon_{e,~-1}^{p-1}\epsilon_{B,~-2}^{(p-2)/4} mJy &\\
       &  &     & (\nu_c < \nu)       \\
\end{array}
\end{equation}

while for a fireball expanding in a wind-like medium we have:

\begin{equation}
\label{eq_wind_opt}
\begin{array}{lccr}
F_\nu & = & 10^{2.3-1.2p}D_{28}^{-2}\nu_{14.6}^{-(p-1)/2}t_d^{-(3p-1)/4} &\\
       &  & \times  E_{53}^{(p+1)/4}A_*\epsilon_{e,~-1}^{p-1} \epsilon_{B,~-4}^{(p+1)/4} mJy &\\
       &  &                                    & (\nu_m < \nu < \nu_c)\\
\end{array}
\end{equation}
       
\begin{equation}
\label{eq_wind_x}
\begin{array}{lccr}
F_\nu  & =&  1.35 \times\left(\frac{17}{72}\right)^{p/4} \times 10^{2.4-0.8p}D_{28}^{-2}\nu_{14.6}^{-p/2}t_d^{-(3p-2)/4} &\\
       &  &  \times E_{53}^{(p+2)/4}\epsilon_{e,~-1}^{p-1}\epsilon_{B,~-2}^{(p-2)/4} mJy &\\
       &  &     & (\nu_c < \nu)       \\
\end{array}
\end{equation}

where $D$, $\nu$, and $t$ are the distance, observation frequency and observation time respectively, and $p$ is the electron distribution power law index; we assume $p=2.2$ for all bursts.
We rewrite these equations in terms of $E_B = \epsilon_B E$ and $E_e = \epsilon_e E$, the energy carried by the magnetic field and relativistic electrons respectively ($E$ is the fireball total energy): 

\begin{equation}
\label{eq_ISM_opt2}
\begin{array}{lccr}
F_\nu & = & 10^{2.1+0.7p}D_{28}^{-2}\nu_{14.6}^{-(p-1)/2}t_d^{-3(p-1)/4} &\\
       &  & \times E_{B,~53}^{(p+1)/4}E_{e,~53}^{p-1}E_{53}^{2-p}(n_*/E_{53})^{1/2} mJy &\\
       &  &                                    & (\nu_m < \nu < \nu_c)\\
\end{array}
\end{equation}
       
\begin{equation}
\label{eq_ISM_x2}
\begin{array}{lccr}
F_\nu  & =&  10^{0.4+0.7p}D_{28}^{-2}\nu_{14.6}^{-p/2}t_d^{-(3p-2)/4} &\\
       &  &  \times E_{B,~53}^{(p-2)/4}E_{e,~53}^{p-1}E_{53}^{2-p} mJy &\\
       &  &     & (\nu_c < \nu)       \\
\end{array}
\end{equation}

\begin{equation}
\label{eq_wind_opt2}
\begin{array}{lccr}
F_\nu & = & 10^{2.3+0.8p}D_{28}^{-2}\nu_{14.6}^{-(p-1)/2}t_d^{-(3p-1)/4} &\\
       &  & \times  E_{B,~53}^{(p+1)/4}E_{e,~53}^{p-1}E_{53}^{2-p}(A_*/E_{53}) mJy &\\
       &  &                                    & (\nu_m < \nu < \nu_c)\\
\end{array}
\end{equation}
       
\begin{equation}
\label{eq_wind_x2}
\begin{array}{lccr}
F_\nu  & =&  1.35 \times\left(\frac{17}{72}\right)^{p/4} \times 10^{0.4+0.7p}D_{28}^{-2}\nu_{14.6}^{-p/2}t_d^{-(3p-2)/4} &\\
       &  &  \times E_{B,~53}^{(p-2)/4}E_{e,~53}^{p-1}E_{53}^{2-p} mJy &\\
       &  &     & (\nu_c < \nu)       \\
\end{array}
\end{equation}

We thus find the optical-to-X-ray flux ratio to be:

\begin{equation}
\label{eq_ratio_ism}
\frac{F_{opt,~ISM}}{F_{X,~ISM}} = 10^{1.7}t_d^{1/4}\nu_{opt,~14.6}^{-(p-1)/2}\nu_{X,~14.6}^{p/2} E_{B,~53}^{3/4}\left(\frac{n_{*}}{E_{53}}\right)^{1/2}
\end{equation}

for a fireball expanding in a ISM and
\begin{equation}
\label{eq_ratio_wind}
\frac{F_{opt,~wind}}{F_{X,~wind}} = \frac{10^{1.9+0.1p}}{1.35}\left(\frac{72}{17}\right)^{p/4} t_d^{-1/4}\nu_{opt,~14.6}^{-(p-1)/2}\nu_{X,~14.6}^{p/2} E_{B,~53}^{3/4}\left(\frac{A_{*}}{E_{53}}\right)
\end{equation}

for a fireball expanding in a wind-like medium.

We extracted from the work of \citet{nar06} the mean optical fluxes at $4.69 \times 10^{14}$ Hz. They are 0.178 mJy and 10.0$\mu$Jy for groups {\it oI} and {\it oII} respectively. The mean X-ray fluxes (at $4.8 \times 10^{17}$ Hz) are 2.10 $\mu$Jy and 0.09 $\mu$Jy for groups {\it I} and {\it II} respectively.

Combining equation \ref{eq_ISM_opt2} with equation \ref{eq_ratio_ism}, and equation \ref{eq_wind_opt2} with equation \ref{eq_ratio_wind}, we obtain the results listed in Table \ref{table_equation}. One can note that the constraints on $E_e$ depend weakly on the medium density (because of the very low exponent of the density parameter). Because $E_e < E$ (or $\epsilon_e < 1$), we can find a constraint on the total fireball energy $E$ (a similar condition is true for $E_B$, but this is not constraining for $E$).

\begin{table*}
\centering
\caption{Constraints on $E_B$, $E_e$, $E$, and the density parameter implied by the clustering observed in X-ray and in optical.\label{table_equation}}
\begin{tabular}{cccccc}
\tableline
\tableline
X-ray   & Optical & Medium & $E_e$      & $E_B$      & $E$ \\
group   & group   & class  & constraint & constraint & constraint\\
\tableline
{\it I} & {\it oI} &  ISM  &$E_{e,53}^{1.2} = 0.04 E_{53}^{1/6} n_0^{1/30}$ & $E_{B,53} = 0.88 \times 10^{-4} (E_{53}/n_0)^{2/3}$ & $0.04 < E_{53}$\\
{\it I} & {\it oI} &  wind &$E_{e,53}^{1.2} = 0.16 E_{53}^{2/15} A_*^{1/15}$& $E_{B,53} = 0.79 \times 10^{-5} (E_{53}/A_*)^{4/3}$ & $0.17 < E_{53}$\\
\tableline
{\it II}& {\it oI} &  ISM  &$E_{e,53}^{1.2} = 0.002 E_{53}^{1/6} n_0^{1/30}$ & $E_{B,53} = 5.84 \times 10^{-3} (E_{53}/n_0)^{2/3}$ & $0.002 < E_{53}$\\
{\it II}& {\it oI} &  wind &$E_{e,53}^{1.2} = 0.006 E_{53}^{2/15} A_*^{1/15}$& $E_{B,53} = 4.95 \times 10^{-4} (E_{53}/A_*)^{4/3}$ & $0.008 < E_{53}$\\
\tableline
{\it II}& {\it oII}&  ISM  &$E_{e,53}^{1.2} = 0.002 E_{53}^{1/6} n_0^{1/30}$ & $E_{B,53} = 1.26 \times 10^{-4} (E_{53}/n_0)^{2/3}$ & $0.002 < E_{53}$\\
{\it II}& {\it oII}&  wind &$E_{e,53}^{1.2} = 0.007 E_{53}^{2/15} A_*^{1/15}$& $E_{B,53} = 1.07 \times 10^{-5} (E_{53}/A_*)^{4/3}$ & $0.009 < E_{53}$\\
\tableline
\end{tabular}
\end{table*}

\subsubsection{The fireball parameters}

Within each group the flux variations around the mean value are small. This may indicate that either $E_B$, $E_e$, $E$, and $n_*$ (or $A_*$, depending on the surrounding medium) are correlated quantities, or are all fixed to a common value. In the former case, because the surrounding medium density is fixed {\it before} the burst occurs, it is this quantity that should be considered as the true variable parameter (and not the energy injected within the fireball). On the other hand, a stellar progenitor produces a wind which affects its surrounding medium \citep[e.g.][]{ram01, che04, dai06}. Thus, the surrounding density is (partly) fixed by the progenitor properties. As these properties should also affect the energy quantity emitted by the fireball (and thus $E_B$, $E_e$, and $E$), we find no surprising that $E_B$, $E_e$, $E$, and $A_*$ or $n_*$ are correlated together.

If $E_B$, $E_e$, $E$, and $n_*$ (or $A_*$) have fixed values, the constraints on the model are strong since this restrict to only three types the possible environments in which a GRB can occur (according to the three observed behaviors). If so, this is a possible explanation of why we do not observe a relativistic outflow associated with each type Ib/c supernovae \citep{sod04}: in fact, normal type Ib/c supernovae can't develop as a GRB.

\subsubsection{The nature of the clustering}

The flux ratio between the two optical groups and the two X-ray groups are similar. \citet{pan02} have investigated the surrounding medium around 10 GRBs (\object{GRB 970508}, \object{GRB 980519}, \object{GRB 990123}, \object{GRB 990510}, \object{GRB 991208}, \object{GRB 991216}, \object{GRB 000301C}, \object{GRB 000418}, \object{GRB 000926}, and \object{GRB 010222}). They found that most of them can be fit with an ISM (all belong to our optical and X-ray bright burst group), while only one burst, \object{GRB 970508}, requires a wind environment (it is an optical and X-ray dim burst). This suggests that {\it group II-oII} bursts are surrounded by a stellar wind while {\it group I-oI} are located within an ISM. Assuming that the only difference between these two groups is due to the medium density profile (and thus that all other parameters have a similar distribution within the two groups), we can explain the difference between the two groups if we set $n_*/E_{53} \sim 1233 A_*^2$ (assuming $\nu_x < \nu_c$). However in such a case we then cannot explain the behavior of {\it group II-oI} bursts. We also note that \citet{pan02} have found that a wind medium can also accurately describe the surrounding environment of \object{GRB 991216} and \object{GRB 010222} (two {\it group I-oI} bursts). Thus, the surrounding medium type  cannot explain the origin of the observed clustering.

In X-ray, the flux is mostly dependent on $E_e$ and $E$ (see eqn. \ref{eq_ISM_x2} and \ref{eq_wind_x2}). If one assumes that $E_{iso}$ is a good estimator of $E$, then since the observed distributions of $E_{iso}$ are similar for the two groups (see Fig. \ref{eiso}), different values of $E$ cannot explain the differences. In turn this implies that  $E_e$ varies within the two groups (see Table \ref{table_equation}). This however again cannot explain the origin of {\it group II-oI}: if the clustering is due only to a difference in the $E_e$ values, then we cannot expect from eqn. \ref{eq_ISM_opt2} and \ref{eq_ISM_x2}, and \ref{eq_wind_opt2} and \ref{eq_wind_x2}, that a dim X-ray afterglow can be bright in optical. The constraints listed in Table \ref{table_equation} indicate that the difference between groups {\it I-oI} and {\it II-oII} on one hand, and group {\it II-oI} on the other hand can be also due to a different value of $E_B$.

We thus propose that the presence of different groups is ascribed to different families of $E_e$ and $E_B$ values:
\begin{itemize}
\item a family of 'magnetized fireball' that produces the group {\it II-oI}. In such a case the fireball transfers only a low fraction of its energy into relativistic electrons. 
\item a family of 'less magnetized fireball' that produces the groups {\it I-oI} and {\it II-oII}. The fraction of total energy going into magnetic fields is roughly one order of magnitude lower than group {\it II-oI}. These two groups can be related to an high and low fraction of energy going in relativistic electrons, respectively. 
\end{itemize}

 Interestingly, the group {\it II-oII} is not very numerous (compared to the other two groups), so that most of GRBs of our sample are then either 'magnetized but not electron-energized' or 'not magnetized but electron-energized'.

\section{Applications}
\label{sec_redshift}

\subsection{The estimation of GRB source redshifts}

The clustering observed in X-ray can be used for distance estimation. To date, most of the redshift measurements made on GRB afterglows were done by optical spectroscopic or photometric observations. However, because of the Lymann alpha cut-off, this method cannot be applied to high redshift events ($z > \sim 5-7$), unless doing IR spectroscopy. Moreover, not all GRBs can be followed in optical \citep[e.g. the so-called dark-bursts,][]{dep05}, either because they are too faint at the time of the observation (or even dark), or because there is no large enough telescope available, or because of the position of the event respective to the Sun. Hence many GRBs have no known distance, stressing the need for intrinsic distance indicators. After the launch of the SWIFT satellite, thanks to its fast re-pointing capabilities, nearly all GRBs have an homogeneous X-ray follow-up. Hence, a redshift measurement method based solely on X-ray observations could be very interesting if one wants to use a large sample of GRB sources for cosmological studies. We propose here to use the reported clustering of GRBs X-ray light curves in three classes to build a redshift estimator. 

The method is based on the redshift needed for a burst to belong to one (or both) of the two groups. The steps are easy to follow: we compute from the X-ray observations (i.e. obtained from SWIFT) the flux at 1 day (observer frame) in the 2-10 keV band, either using a mean spectral index of $\sim 1.2$, or the exact one obtained from X-ray spectral fitting. We note that even if the spectral index is not known, this has little influence on the actual flux at one day (within reasonable limits). Table \ref{table_estimate}, which provides the redshift needed to comply with the relation for both groups for a given observed flux, is directly usable to estimate the redshift.

\begin{figure}
  \includegraphics[width=8cm]{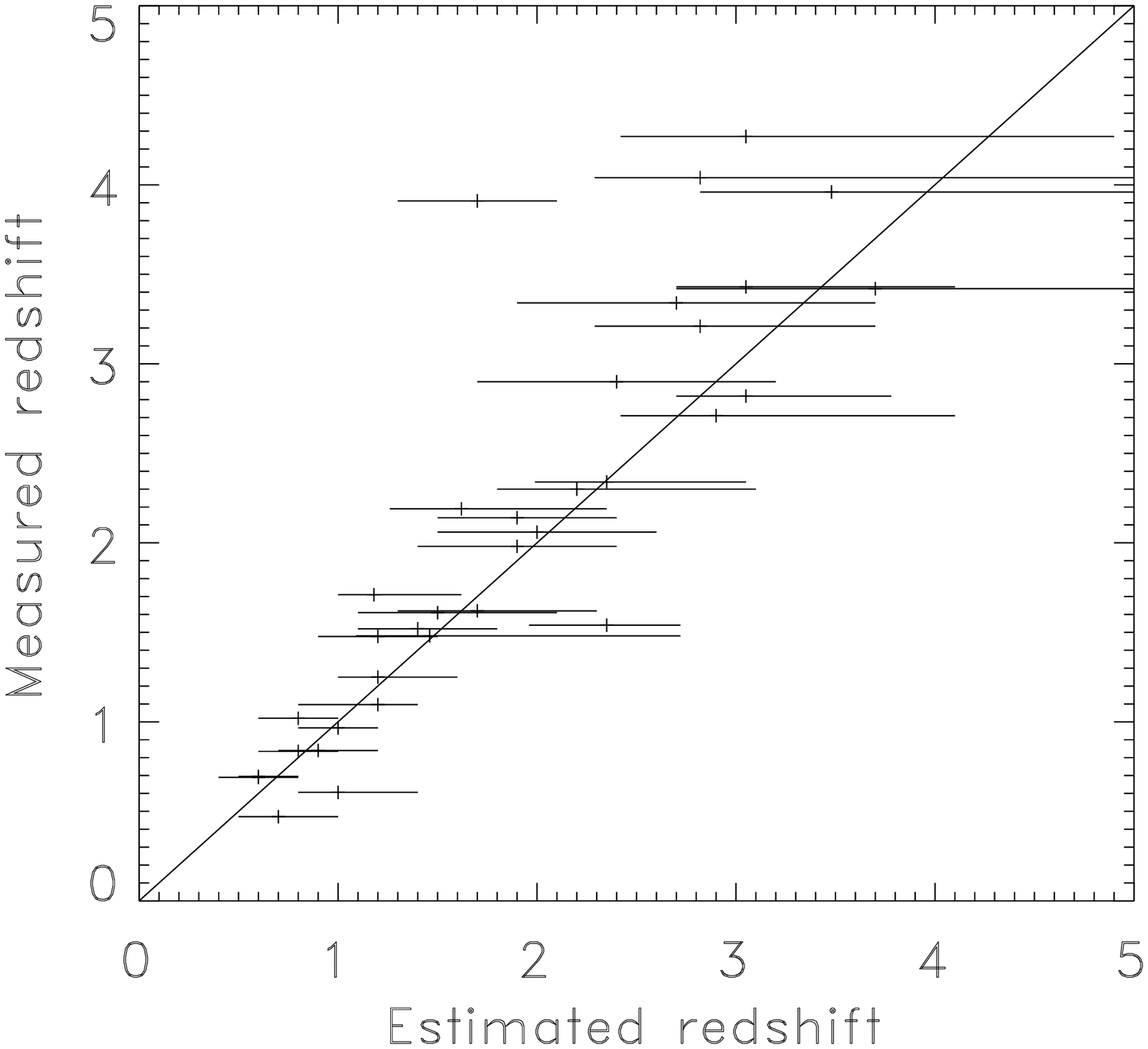}
  \caption{\label{test_z} Comparison of the estimated redshift versus the measured one for the bursts defining our sample. The solid line corresponds to equality .}
\end{figure}

We calibrated this method by deriving the estimated redshift for the bursts of our sample (for which the group is known), and comparing this value with the measured redshift. The results are displayed in Fig. \ref{test_z}. The estimated redshift agrees with the measured one for most of the bursts. The only discrepancies arises at low redshift. In Sect. \ref{sec_discu} we already noted that {\it group III} bursts are located at small distances (z$<$0.5) and do not follow the same relation as groups I and II: as a conservative approach we prefer to restrict the validity of our method to source located at redshifts larger than 0.5. We list in Table \ref{table_estimate} the observed flux one day after the burst (observer frame) and the associated redshift needed for a burst to comply with the observed clustering for group I and II sources. Because of the limitation to redshifts larger than 0.5, all bursts brighter than $\sim 2 \times 10^{-12}$ erg s$^{-1}$ cm$^{-2}$ belong to {\it group I}. However, the belonging group is not correlated with easily observable quantities (absorption, peak energy, isotropic energy, spectral or decay indexes), so it can be fixed only through broad-band modeling. In fact this leads to two redshift estimates. If the bursts has an optical afterglow there is a possible way to decide which of the estimate is valid: a burst with flux of $\sim 2 \times 10^{-12}$ erg s$^{-1}$ cm$^{-2}$ which belongs to {\it group I} has a redshift of $\sim 2$. At that distance, the Lymann $\alpha$ cut-off appears to be at 360nm, in the UVOT observation bands. Hence, if a GRB afterglow is observed by UVOT (or in the B band for ground based telescopes) then it cannot belong to group I, and the ambiguity on the redshift determination is cleared. 

\begin{table}

\caption{Flux to redshift conversion. This table should be used as an estimate of the redshift for bursts observed by any X-ray observatory. The flux is given in the 2-10 keV band at 1 day after the burst (observer frame). The redshift has been calculated for an energy index of 1.2, the uncertainty is 30\%.\label{table_estimate}}
\centering                          
\begin{tabular}{c c c }        
\tableline\tableline                 
Flux & Group I redshift & group II redshift \\    
(erg s$^{-1}$ cm$^{-2}$)&&                  \\
\tableline                        
$1 \times 10^{-14}$ & ---    &  4.43  \\
$2 \times 10^{-14}$ & ---    &  3.28  \\
$3 \times 10^{-14}$ & ---    &  2.72  \\
$4 \times 10^{-14}$ & ---    &  2.35  \\
$5 \times 10^{-14}$ & ---    &  2.12  \\
$6 \times 10^{-14}$ & 7.80   &  1.96  \\
$7 \times 10^{-14}$ & 7.09   &  1.83  \\
$8 \times 10^{-14}$ & 6.38   &  1.75  \\
$9 \times 10^{-14}$ & 6.05   &  1.68  \\
$1 \times 10^{-13}$ & 5.78   &  1.62  \\
$2 \times 10^{-13}$ & 4.10   &  1.26  \\
$3 \times 10^{-13}$ & 3.48   &  1.09  \\
$4 \times 10^{-13}$ & 3.05   &  0.96  \\
$5 \times 10^{-13}$ & 2.82   &  0.89  \\
$6 \times 10^{-13}$ & 2.59   &  0.82  \\
$7 \times 10^{-13}$ & 2.42   &  0.77  \\
$8 \times 10^{-13}$ & 2.29   &  0.73  \\
$9 \times 10^{-13}$ & 2.10   &  0.69  \\
$1 \times 10^{-12}$ & 1.99   &  0.66  \\
$2 \times 10^{-12}$ & 1.50   &  0.50  \\
$3 \times 10^{-12}$ & 1.29   &  ---  \\
$4 \times 10^{-12}$ & 1.16   &  ---  \\
$5 \times 10^{-12}$ & 1.06   &  ---  \\
$6 \times 10^{-12}$ & 0.99   &  ---  \\
$7 \times 10^{-12}$ & 0.94   &  ---  \\
$8 \times 10^{-12}$ & 0.89   & ---   \\
$9 \times 10^{-12}$ & 0.83   & ---   \\
$1 \times 10^{-11}$ & 0.79   &  ---  \\
$2 \times 10^{-11}$ & 0.56   &  ---  \\
$3 \times 10^{-11}$ & 0.50   &  ---  \\

\tableline                                   
\end{tabular}
\end{table}

\subsection{Estimated redshift of pre-SWIFT bursts with unknown distance}
\label{toto}

\begin{table}
\caption{Redshift estimates derived from the relation using either the group I and II hypotheses for pre-SWIFT bursts without known distance.\label{table2}}
\centering                          
\begin{tabular}{ccc}
\tableline
\tableline
GRB  name  &  Redshift           & estimate     \\
           &  group I            & group II     \\
\tableline
\object{GRB 980329} & 4.2 $\pm$ 1.2       & 1.2 $\pm$ 0.2\\
\object{GRB 980519} & 3.8 $\pm$ 0.7       & 1.4 $\pm$ 0.2\\
\object{GRB 990704} & 3.5 $\pm$ 0.9       & 1.3 $\pm$ 0.3\\
\object{GRB 990806} & 4.7$^{+1.6}_{-0.7}$ & 1.6 $\pm$ 0.3\\
\object{GRB 001109} & 2.3 $\pm$ 0.7       & 0.8 $\pm$ 0.2\\
\object{GRB 001025A}& 5.8 $\pm$ 1.8       & 2.2 $\pm$ 0.4\\
\object{GRB 020322} & 5.0 $\pm$ 1.5       & 1.5 $\pm$ 0.3\\
\object{GRB 020410} & 0.5 $\pm$ 0.4       & $<0.1$       \\
\object{GRB 040106} & 3.4 $\pm$ 0.5       & 1.0 $\pm$ 0.2\\
\object{GRB 040223} & 5.5$^{+2.0}_{1.2}$  & 1.7 $\pm$ 0.2\\
\object{GRB 040827} & 8.0 $\pm$ 2.0       & 1.9 $\pm$ 0.3\\
\tableline
\end{tabular}
\end{table}

We retrieved from \citet{gen05b} and  \citet{dep06} the light curves of several bursts with good temporal sampling and spectral informations (table \ref{table2}), and estimated their redshift assuming they belong either to {\it group I} or to {\it group II}. For only one burst we compute an associated redshift lower than 0.5 (\object{GRB 020410}), and thus we do not consider it in the following discussion. 

The estimated redshift of \object{GRB 980519} is either $3.8\pm0.7$ or 1.4 $\pm$ 0.2. This burst was observed in U band \cite{jau01,kan06}. As the Lymann $\alpha$ cut-off cross the U band at $z \sim 2.8$, this burst should not have been observed in the high distance hypothesis. Thus, U observations can indeed solve the problem of group classification, and we propose a redshift measurement for \object{GRB 980519} of 1.4 $\pm$ 0.2 based on X-ray observations. For the same reason we argue that \object{GRB 040827} is at $z = 1.9 \pm 0.3$ and not $z = 8 \pm 2$, since an optical afterglow has been observed in optical \citep{mel04}.

\object{GRB 001025A} was observed by XMM-Newton but no optical afterglow was detected. This event was thus classified as a dark burst \citep{ped05}. We note that this one has a large redshift value, most of all if it belong to {\it group I}. In such a case, with a calculated redshift of $5.8 \pm 0.8$, we can explain the classification as dark burst by the Lyman alpha cut-off at $\sim 6200 \AA$. However, since we cannot draw conclusions from the absence of detection at optical wavelengths, and as dark bursts can also be intrinsically fainter \citep{dep05}, we cannot exclude the hypothesis that this event is located at $z = 2.2 \pm 0.4$, hence belonging to {\it group II}.

Finally, the mean redshift for SWIFT bursts is 2.3, while the pre-SWIFT mean redshift was 1.2 \citep{gru07}. Lets assume that {\it all} bursts listed in Table \ref{table2} are {\it group II} bursts, a very conservative hypothesis, the mean redshift, in this case, is 1.5 : this is already larger than the pre-SWIFT mean redshift. Since it is reasonable to consider that at least some of these events belong to {\it group I}, this should be considered as a lower limit. Before SWIFT, the X-ray/optical follow-up was late, and only the brightest events were observed. As more distant events will appear fainter, this introduced a bias against distant events. Indeed, the fact that these bursts without known redshift are slightly more distant indicates that the pre-SWIFT redshift distribution was biased toward low redshift due to selection effects, and that SWIFT cleared this bias thanks to its fast follow-up.

\begin{table}
\caption{Redshift estimates derived from the relation using either the group I and II hypotheses for SWIFT bursts with known distance.\label{table_sup}}
\centering                          
\begin{tabular}{cccc}
\tableline
\tableline
GRB  name           &  Group I  & Group II & Measured   \\
                    &  estimate & estimate & redshift  \\
\tableline
\object{GRB 070529} &  7.8      &  1.96    & 2.44 \\
\object{GRB 070611} &  6.05     &  1.68    & 2.04 \\
\object{GRB 070721B}&  ---      &  3.28    & 3.62 \\
\tableline
\end{tabular}
\end{table}

\section{Conclusions}

We have investigated the clustering of afterglow light curves observed previously (papers I and II) in X-ray and in optical from BeppoSAX, XMM-Newton and Chandra data, using the newly available, well-monitored SWIFT light curves. Adding SWIFT bursts to the previous sample reported in paper II, we still confirm our previous findings. On a sample of 61 events the X-ray light curves cluster in two groups, with a significance larger than 6 $\sigma$. \citet{wil07} describe the X-ray light curve as a sum of two components, the first one being related to the prompt emission: we found that it is the second component that clusters in luminosity. This finding is also supported by the fact that the clustering is not related to the properties of the prompt emission.

A similar clustering was observed also at optical wavelengths. We compared the classification within each group in X-ray and in optical, and found three classes: {\it bright} optical and X-ray afterglows, {\it dim} ones, and optically {\it bright}-X-ray {\it dim} ones. We propose that this clustering is related to the fireball properties and the surrounding medium density, which can be either constant or correlated. We used the mean X-ray and optical fluxes of each group to put constraints on the values of $\epsilon_B$ and $\epsilon_{\bf e}$, and expressed them in terms of the total fireball energy and the surrounding medium density. We stress that the extension of this work at low frequency (radio, sub-millimeter and infrared) may help solving the exact origin of the clustering by strongly constraining the medium density and the position of the synchrotron self absorption and injection frequencies. 

Using the observed X-ray afterglow properties of GRBs, we propose a new, simple, method for the determination of the source redshift based on X-ray data; optical photometry in U and V bands may help to clear the degeneracy between the two estimates found. This method has been established on a sample of bursts with known redshift, and we find an excellent correlation. We apply this method to a sample of GRB source of unknown redshift. We propose an estimation of the redshift for \object{GRB 980519} (1.4 $\pm$ 0.2) and for \object{GRB 040827} ($1.9 \pm 0.3$).

While we were building this distance indicator and writing this paper, several GRB were detected by SWIFT and their distance derived using optical spectroscopy. We thus have a sample of bursts which allows to test this correlation independently. To simulate actual conditions for an unknown event we used the SWIFT count rate light curve available from their web site \citep{eva07}, and we apply a mean spectral index of 1.2 for the count-to-flux conversion. The results are indicated in Table \ref{table_sup}, indicating an agreement between 10.3 and 24.5 \% \citep[in accordance with the error estimated in \ref{toto}, we estimate far superior than other methods e.g.][]{att05}.

\acknowledgements
BG acknowledge support from COFIN grant 2005025417, and from the Centre Nationnal d'Etudes Spatiales. We are pleased to thank Alessandra Corsi, Giulia Stratta and an anonymous referee for very useful and constructive discussions. This work made use of data supplied by the UK Swift Science Data Centre at the University of Leicester.

{\it Facilities:} \facility{Swift}, \facility{CXO}, \facility{BeppoSAX}, \facility{XMM-Newton}.


\begin{thebibliography}{}

\bibitem[Amati et al. (2002)]{ama02} Amati, L., Frontera, F., Tavani, M., et al., 2002, \aap, 390, 81
\bibitem[Amati (2006)]{ama06} Amati, L., 2006, \mnras, 372, 233
\bibitem[Atteia (2003)]{att05} Atteia, J.L., 2003, \aap, 407, L1
\bibitem[Berger et al. (2005)]{ber05} Berger, E., Kulkarni, S.R., Fox, D.B., et al., 2005, \apj, 634, 501
\bibitem[Bloom et al. (2006)]{blo06} Bloom, J. S., Foley, R. J., Koceveki, D., \& Perley, D., 2006, GCN \#5217
\bibitem[Bo\"er \& Gendre (2000)]{boe00} Bo\"er, M. \& Gendre, B. 2000, A\&A, 361, L21
\bibitem[Bo\"er et al. (2006)]{boe06} Bo\"er, M., Atteia, J.L., Damerdji, Y.,  et al., 2006, \apj, 638, L71
\bibitem[Chattopadhyay et al. (2007)]{cha07} Chattopadhyay, T., Misra, R., Chattopadhyay, A.K., \& Naskar, M., 2007, \apj, 667, 1017
\bibitem[Chevalier et al. (2004)]{che04} Chevalier, R.A., Li, Z.Y., \& Fransson, C., 2004, \apj, 606, 369
\bibitem[Eldridge et al. (2006)]{dai06} Eldridge, J. J., Genet, F., Daigne, F., \& Mochkovitch, R., 2006, MNRAS, 367, 186
\bibitem[Evans et al. (2007)]{eva07} Evans, P. A., Beardmore, A. P., Page, K. L., et al., 2007, \aap, 469, 379
\bibitem[Gendre \& Bo\"er (2005)]{gen05} Gendre, B., Bo\"er, M., 2005, A\&A , 430, 465
\bibitem[Gendre et al.  (2006)]{gen05b} Gendre,  B., Corsi, A., \&  Piro,  L.,   2005, \aap, 455, 803
\bibitem[Gendre \& Bo\"er (2006)]{gen06} Gendre, B., \& Bo\"er, M., 2006 AIPC, 836, 558
\bibitem[Gendre et al. (2007)]{gen07} Gendre, B., Galli, A., Corsi, A., et al., 2007, \aap, 462, 565
\bibitem[Ghirlanda et al. (2004)]{ghi03} Ghirlanda, G., Ghisellini, G., \& Lazzati, D., 2004, \apj, 616, 331
\bibitem[Grupe et al. (2007)]{gru07} Grupe, D., Nousek, J., van den Berk, D.E., et al., 2007, \aj, 133, 2216
\bibitem[Hjorth et al. (2003)]{hjo03} Hjorth, J., Sollerman, J., M\o ller, P., et al. 2003, Nature, 423, 847
\bibitem[Jakobsson et al. (2006)]{jac06} Jakobsson, P., Levan, A., Fynbo, J. P. U., et al., 2006, \aap, 447, 897
\bibitem[Jaunsen et al. (2001)]{jau01} Jaunsen, A.O., Hjorth, J., Bj\"ornsson, G., et al., 2001, \apj, 546, 127
\bibitem[Kann et al. (2006)]{kan06} Kann, D.A, Klose, S., \& Zeh, A., 2006, \apj, 641, 993
\bibitem[Kouveliotou et al. (2004)]{kou04} Kouveliotou, C., Woosley, S.E., Patel, A., et al., 2004, \apj, 608, 872
\bibitem[Lamb \& Reichart (2000)]{lam00} Lamb \& Reichart, 2000, \apj, 536, L1
\bibitem[Liang \& Zhang (2006)]{lia06} Liang, E. \& Zhang, B., 2006, \apj, 638, L67
\bibitem[Malesani et al (2004)]{mel04} Malesani, D., D'Avanzo, P., Melandri, A., et al., 2004, GCN \# 2685
\bibitem[Metzger et al. (1997)]{met97} Metzger, M.R., Djorgovski, S.G., Kulkarni, S.R., et al., 1997, Nature, 387, 879
\bibitem[Meszaros (2006)]{mes06} Meszaros, P, 2006, Rep. Prog. Phys., 69, 2259 
\bibitem[Nardini et al. (2006)]{nar06} Nardini, M., Ghisellini, G., Ghirlanda, G., et al., 2006, \aap, 451, 821
\bibitem[Nousek et al. (2006)]{nou05} Nousek, J. A., Kouveliotou, C., Grupe, D., et al., 2006, \apj, 642, 389
\bibitem[O'Brien et al. (2006)]{obr06} O'Brien, P.T., Willingale, R., Osborne, J., et al., 2006, \apj, 647, 1213
\bibitem[Panaitescu \& Kumar (2000)]{pan00} Panaitescu, A., \& Kumar, P., 2000, \apj, 543, 66
\bibitem[Panaitescu \& Kumar (2002)]{pan02} Panaitescu, A., \& Kumar, P., \apj, 571, 779, 2002
\bibitem[De Pasquale et al. (2003)]{dep05} De Pasquale, M., Piro, L., Perna, R., et al., 2003, \apj, 592, 1018
\bibitem[De Pasquale et al. (2006)]{dep06} De Pasquale, M., Piro, L., Gendre, B., et al.,  2005, \aap, 455, 813
\bibitem[Ramirez-Ruiz et al. (2001)]{ram01} Ramirez-Ruiz, E., Dray, L.M., Madau, P., Tout, C.A., 2001, \mnras, 327, 829
\bibitem[Pedersen et al. (2006)]{ped05} Pedersen, K., Hurley, K., Hjorth, J., et al., 2006, \apj, 636, 381
\bibitem[Soderberg et al. (2004)]{sod04} Soderberg, A.M., Frail, D.A., \& Wieringa, M.H., 2004, \apj, 607, L13
\bibitem[Stanek et al. (2003)]{sta03} Stanek, K. Z., Matheson, T., Garnavich, P. M., et al., 2003, \apj, 591, L17
\bibitem[Watson et al. (2006)]{wat05} Watson, D., Reeves, J.N., Hjorth, J., et al., 2006, \apj, 637, L69
\bibitem[Willingale et al. (2007)]{wil07} Willingale, R., O'Brien, P. T., Osborne, J. P., et al., 2007, \apj, 662, 1093
\bibitem[Zhang et al. (2006)]{zhe06} Zhang, B., Fan, Y.Z., Dyks, J., et al., 2006, \apj, 642, 354








\end{thebibliography}
\end{document}